\documentclass[10pt]{article}
\usepackage{amsmath}
\usepackage{amssymb}
\usepackage{graphicx}

\begin{document}

\setcounter{figure}{0}
\setcounter{table}{0}
\setcounter{footnote}{0}
\setcounter{equation}{0}

\vspace*{0.5cm}

\noindent {\Large RADIO QUASARS AND THE LINK WITH GAIA}
\vspace*{0.7cm}

\noindent\hspace*{1.5cm} J. ROLAND$^1$ \\
\noindent\hspace*{1.5cm} $^1$ Institut d'Astrophysique de Paris\\
\noindent\hspace*{1.5cm} 98 Bis Bd Arago, 75014 Paris, France\\
\noindent\hspace*{1.5cm} e-mail: jacques.roland.7@gmail.com\\

\vspace*{0.5cm}

\noindent {\large ABSTRACT.} 
Modeling VLBI ejections of nuclei of extragalactic radio sources, indicates that 
their nuclei contain a binary black hole system. One can derive the distance and the 
positions of the two black holes in the plane of the sky. We can also 
use the RMS of the time series of the ICRF2 survey to obtain an estimate of the structure 
and the size of the nuclei. We will discuss the possible problems to link VLBI 
observations and GAIA optical observations of radio quasars if they contain a binary 
black hole system. 

\vspace*{1cm}

\noindent {\large 1. STRUCTURE OF COMPACT RADIO SOURCES MODELING VLBI EJECTIONS}

\smallskip

VLBI observations of compact radio sources show that the ejection 
of VLBI components does not follow a straight line but undulates. 
These observations suggests a precession of the accretion disk. 
To explain the precession of the accretion disk, we will assume that 
the nucleus of radio sources contains a binary black hole system 
(BBH system) .

A BBH system produces 2 main perturbations of the VLBI ejection due to:
\begin{enumerate}
    \item the precession of the accretion disk and
    \item the motion of the two black holes around the gravity center 
    of the BBH system.
\end{enumerate}

\begin{figure}[ht]
\begin{center}
\includegraphics[scale=0.6]{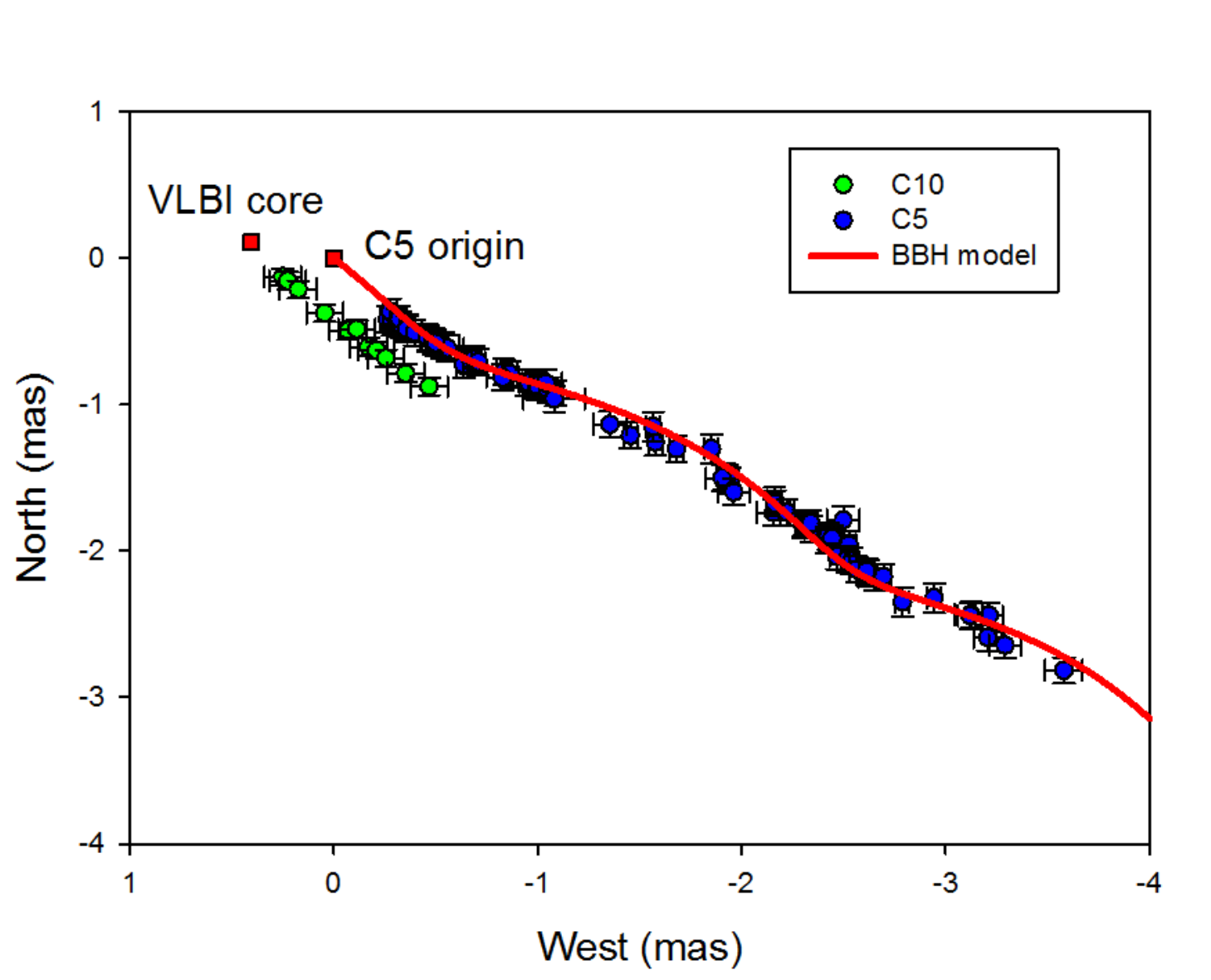}
\caption{Using the MOJAVE data (Lister \& al 2009), we plot the trajectories of C5 and C10. 
Component C10 is ejected by the VLBI core and component C5 is ejected 
with a large offset from the VLBI core. Components C5 and C10 follow two different 
trajectories and are ejected from different origins, indicating that the 
nucleus of 3C 279 contains a BBH system (Roland \& al. 2013).}
\end{center}
\end{figure}

The presence of a BBH system, induces several consequences, which are:
\begin{enumerate}
    \item the 2 black holes can have accretion disks with different 
    angles with the plane of rotation of the BBH system and can eject 
    VLBI components; in that case we will observe two different families 
    of trajectories, a good example of a source showing 2 families of trajectories 
    is 3C 279 (see Figure 1),
    \item if the VLBI core is associated with one black hole and if the 
    ejection of the VLBI component comes from the second black hole, 
    there will be an offset between the VLBI core and the origin of the 
    ejection of the VLBI component; this offset will correspond the radius 
    of the BBH system.
\end{enumerate}

%\begin{figure}[ht]
%\begin{center}
%\includegraphics[scale=0.5]{1_3C279_C5+C10_Trajectories.pdf}
%\caption{Figure1 caption}
%\end{center}
%\end{figure}

We model the ejection of the VLBI component using a geometrical model 
taking into account the two main perturbations due to the BBH system.
We determine the free parameters of the model comparing the observed 
coordinates of the VLBI component with the calculated coordinates of the 
model.

Modeling the ejection of VLBI components using a BBH system has been 
developed in previous articles, Britzen \& al. 2001 modeled 
0420-014, Lobanov \& Roland 2005 modeled 3C 345, 
Roland \& al. 2008 modeled 1803+784, and Roland \& al. 2013 modeled 3C 279 
and 1823+568.

Results concerning 3C 279 are shown in Figure 1.

\vspace*{0.7cm}

\noindent {\large 2. STRUCTURE USING THE RMS OF THE TIME SERIES OF THE ICRF2 SURVEY}

\smallskip

The ICRF2 Survey (International Celestial Reference Frame) has been obtained 
using about 6.5 millions of VLBI observations of about 3400 radio sources 
(Fey \& al. 2010).

Important information concerning the structure of the nucleus can be obtained using 
the RMS of the time series of the ICRF2 survey (Lambert 2013 and http://ivsopar.obspm.fr/).
To begin, let us take the example of the source 1803+784.
The RMS of the time series of 1803+784 are presented in Figure 2.

\begin{figure}[ht]
\begin{center}
\includegraphics[scale=0.3]{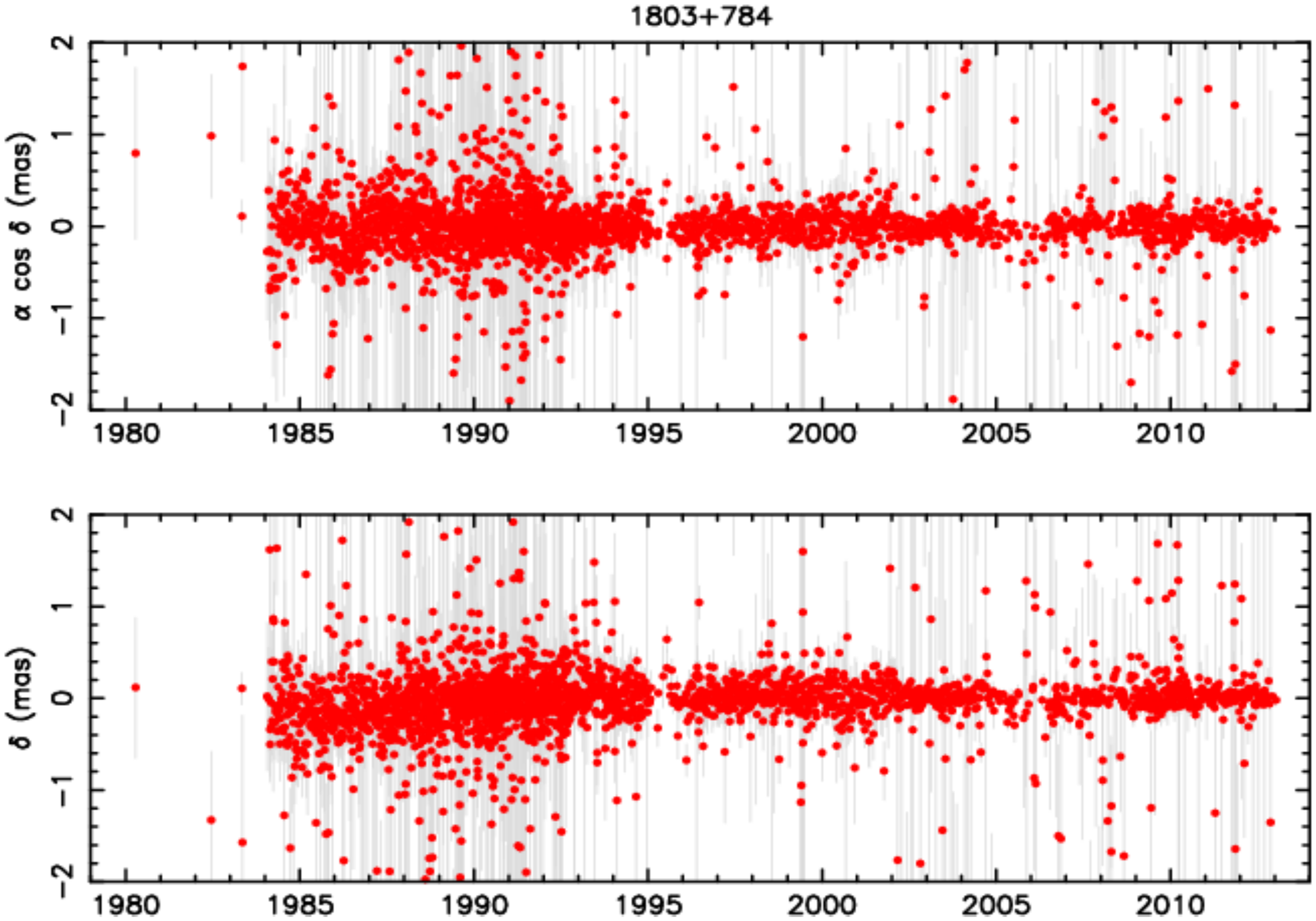}
\caption{The RMS of the time series for the coordinates of 1803+784 are $\approx 0.12$ $mas$ and 
$\approx 0.12$ $mas$. It has been shown by Roland \& al 2008 that the nucleus of 1803+784 
contains a BBH system of size $\approx 0.10$ $mas$.}
\end{center}
\end{figure}

More generally, using the results obtained from the modeling of VLBI ejections, 
it can been shown that the RMS of the time series are correlated with the structure of the nucleus (see Table 1). 
Indeed, the RMS of the time series is always larger than the size of the BBH system deduced from modeling 
VLBI ejections.

\begin{center}
Table 1 : Structures of compact sources and RMS of the time series\medskip%
\end{center}
\begin{center}
\begin{tabular}
[c]{c||c|c}\hline
Source                 & Structure                                                           & RMS time series        \\\hline
PKS 0420-014           & BBH system : $R_{bin} \approx 0.12$ $mas$  (Britzen et al 2001)     & $0.32$ * $0.47$        \\\hline
3C 345                 & BBH system :                           (Lobanov \& Roland 2005)     & $0.71$ * $0.69$        \\\hline
S5 1803+784            & BBH system : $R_{bin} \approx 0.10$ $mas$  (Roland et al 2008)      & $0.12$ * $0.12$        \\\hline
1823+568               & BBH system : $R_{bin} \approx 0.06$ $mas$  (Roland et al 2013)      & $0.16$ * $0.21$        \\\hline
3C 279                 & BBH system : $R_{bin} \approx 0.42$ $mas$  (Roland et al 2013)      & $0.90$ * $1.11$        \\\hline
PKS 1741-03            & BBH system : $R_{bin} \approx 0.18$ $mas$  (Work in progress)       & $0.20$ * $0.23$        \\\hline
1928+738               & BBH system : $R_{bin} \leq    0.23$ $mas$  (Work in progress)       & $0.22$ * $0.35$        \\\hline
3C 345                 & 3 BH or 2 BBH systems                      (Work in progress)       & $0.71$ * $0.69$        \\\hline
\end{tabular}
\end{center}

The ICRF survey has been done at 8 GHz and the smallest RMS of the time series found are $\approx 0.1$ $mas$ at this frequency.
The IRCF survey is now going to be done at 22 GHz and 32 GHz and one can expect to reach for point source sources 
RMS of the time series of 0.03 $mas$. If the smallest RMS of the time series at these frequencies are, say, $\approx 0.08$ $mas$, 
this will mean that the sources are not point sources but contain BBH systems which sizes are $R_{bin} \approx 0.07$ $mas$.

So we can use the RMS of the time series to look for compact radio sources.

%\begin{table}[h]
%\begin{center}
%
%
%
%\caption{Table1 caption}
%\end{center}
%\end{table}

\vspace*{0.7cm}

\noindent {\large 3. LINK BETWEEN VLBI OBSERVATIONS AND GAIA}

\smallskip

GAIA will be able to provide a very precise position but has a relatively 
low resolution (compared to VLBI). For point sources which magnitude is 
$m_{v} \approx 15$ the precision of the position will be $\approx 0.02$ $mas$, 
but for point sources which magnitude is $m_{v} \approx 18$ the precision 
of the position will be $\approx 0.10$ $mas$.

The optical emission from a radio quasar can be due to
\begin{itemize}
	\item the non thermal core (optical emission of the ultra relativistic 
	$e^{-}e^{+}$ ejected relativistically),
	\item the black body radiation of the central parts of the accretion disk,
	\item broad line region and
	\item the stars.
\end{itemize}

The optical emission of radio quasars is dominated by the non thermal emission 
(synchrotron and/or inverse Compton emissions). This result is indicated by 
the power law distribution of the spectrum from the radio to the Xray emission 
and the linear polarization of the emission (see the spectrum of 3C 273 shown 
in Figure 3).

\begin{figure}[ht]
\begin{center}
\includegraphics[scale=0.45]{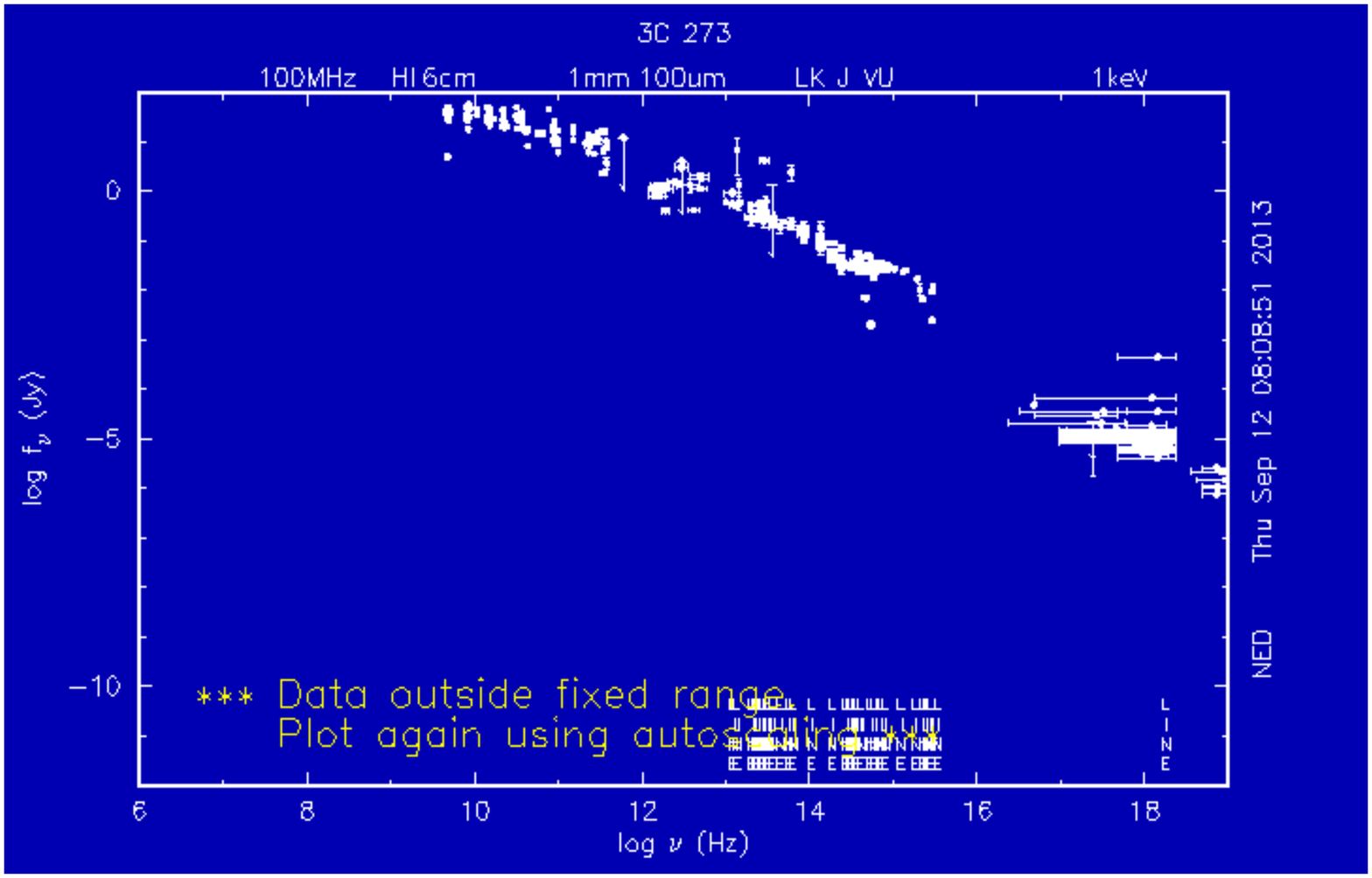}
\caption{The spectrum of 3C 273. The spectrum shows a power law distribution 
between the radio to X and $\gamma$ rays, indicating a non thermal origin. The 
radiation is linearly polarized. This caption is from the NASA/IPAC Extragalactic Data Base 
(http://ned.ipac.caltech.edu/)}
\end{center}
\end{figure}

Due to opacity effect, the optical core and the radio core positions are 
not the same. However, if the inclination angle of the 
source is very small, i.e. $i_{o} \leq 3^{o}$, opacity effect will be small.
The position of the black hole emitting the VLBI jet is not the same that 
the positions of the optical core and the radio core.

If the nucleus of the radio quasar contains a BBH system and if the two black 
holes are active, three different cases can happen:
\begin{enumerate}
	\item the radio core and the optical core are associated with the same BH, then
the distance between the radio core and the optical core depends on the opacity
effect which will be small if the inclination angle is small,
  \item the radio core and the optical core are associated with different black holes, then
the distance between the radio core and the optical core is more or less the size
of the BBH system (corrected by the possible opacity effect), and
  \item the two black holes are emitting in the optical, then
GAIA will provide a mean position between the two optical cores ! This position will be different
from the positions of the two radio cores.
\end{enumerate}
as quasars are strongly and rapidly variables, during the 5 years of 
observations of GAIA, the 3 different cases can happen for a given source !

\vspace*{0.7cm}

\noindent {\large 4. CONCLUSION}

\smallskip

To link, with a precision of $\leq 150$ $\mu as$, the Local Reference Frame obtained by GAIA 
and the Reference Frame provided by distant radio quasars, one has to use 
radio quasars which magnitude is $m_{v} \leq 18$ and which are a priori the most compact.\\

Ideally, one has to define a sample of at least 10 radio quasars characterized by:
\begin{itemize}
	\item $m_{v} \leq 18$,
	\item a RMS of the time series $RMS \leq 200$ $\mu as$ and
	\item a declination $ -90^{o} \leq \delta \leq 90^{o}$,
\end{itemize}
which is currently not possible. Indeed, if we look for sources which have a RMS of the time 
series $RMS \leq 200$ $\mu as$, we find only 10 sources with a declination $\delta > 0^{o}$. To 
obtain sources with $\delta < 0^{o}$, one has to look to sources with a RMS of the time series 
$RMS \leq 500$ $\mu as$ ! This is due to the lack of VLBI observations in the south hemisphere. 

Modeling VLBI ejection using sources from this sample can be done if each VLBI component has been 
observed at least 20 times and if the component can be followed on a path long enough. It has 
the advantage to provide the size of the BBH system, the positions of the two black holes 
and the inclination of the radio source.

During the five years of GAIA observations it will be crucial to improve the number 
and the quality (UV coverage) of VLBI observations of radio sources with negative declinations 
in order to model the VLBI ejections and to reduce significantly the RMS of the time series. 

\vspace*{0.7cm}

\noindent {\large 5. REFERENCES}
% Please type the reference as follows
% Name Initial, year, "title", journal, vol. , pp. x-x.
%
% Examples:
%
% Author1, N., Author2, N., 2000, ``Title of the paper'', 
% \aa 111, pp. 111--222.
%
% Author2, N., Author3, N., 2003, ``Title of the paper'',
% \jgr (Solid Earth), 111(B5), doi: 10.1000/2002JB001111.
%
% PLEASE DO NOT USE ANY SPECIAL FONTS 
% (no italics, no boldface, etc.)
%
{

\leftskip=5mm
\parindent=-5mm

\smallskip

Britzen, S., Roland, J., Laskar, J., 2001,  "On the origin of compact radio sources. 
The binary black hole model applied to the gamma-bright quasar PKS 0420-014", A\&A, 374, 784

Fey, A.L., Gordon, D.G., Jacobs, C.S. \& al., 2009, "International Earth Rotation and 
Reference Systems Service (IERS) Technical Note 35 " in Bundesamts f\"{u}r Kartographie 
und Geod\"{a}sie, Ed. Frankfurt am Main

Lambert, S., 2013,  "Time stability of the ICRF2 axes", A\&A, 553, 122

Lister, M.~L., Aller, H.~D., Aller, M.~F., \& al, 2009, 
"MOJAVE: Monitoring of Jets in Active Galactic Nuclei with VLBA Experiments. V. Multi-Epoch VLBA Images", 
AJ, 137, 3718-3729

Lobanov, A.P., Roland, J., 2005, "A supermassive binary black hole in the quasar 3C 345", 
A\&A, 431, 831

Roland, J., Britzen, S., Kudryavtseva, N.A., Witzel, A., Karouzos, M., 2008,  
"Modeling nuclei of radio galaxies from VLBI radio observations. Application to the BL Lac Object S5 1803+784", 
A\&A, 483, 125

Roland, J., Britzen, S., Caproni, A., Fromm, C., Gl{\"u}ck, C., Zensus, A., 2013, 
"Binary black holes in nuclei of extragalactic radio sources", A\&A, 557, 85

}

\end{document}